\documentclass[%
 article,
 amsmath,amssymb,
 reprint,%
 longbibliography,
]{revtex4-1}

\usepackage{graphicx}
\usepackage{dcolumn}
\usepackage{bm}

\usepackage[utf8]{inputenc}
\usepackage[T1]{fontenc}
\usepackage{mathptmx}
\usepackage{color}
\usepackage{amsmath}
\usepackage[ruled,lined]{algorithm2e}

\usepackage[page]{appendix}

\usepackage{caption}
\usepackage{subcaption}

\usepackage{graphicx}

\usepackage{tikz}
\usepackage{pgfbaselayers}
\usetikzlibrary{quantikz}
\usepackage{url}

\usepackage[textwidth=7in,top=1in,bottom=1.1in,tmargin=0.4in]{geometry}
\usepackage{fancyhdr}
\usepackage{multirow}
\usepackage{booktabs}
\newcommand{\snTwo}{S$_{\mathrm{N}}$2~}

\begin{document}

\pagestyle{fancy}
\fancyhf{}
\cfoot{\thepage}

\title{Enhancing the Electron Pair Approximation with Measurements on Trapped Ion Quantum Computers}

\author{Luning Zhao}
\email{zhao@ionq.co}
\affiliation{
IonQ Inc, College Park, MD, 20740, USA
}

\author{Qingfeng Wang}
\affiliation{
Chemical Physics Program and Institute for Physical Science and Technology, University of Maryland, College Park, MD, 20742, USA
}

\author{Joshua Goings}
\affiliation{
IonQ Inc, College Park, MD, 20740, USA
}

\author{Kyujin Shin}
\email{shinkj@hyundai.com}
\affiliation{
 Materials Research \& Engineering Center, CTO Division, Hyundai Motor Company, Uiwang 16082, Republic of Korea
}

\author{Woomin Kyoung}
\affiliation{ 
  Materials Research \& Engineering Center, CTO Division, Hyundai Motor Company, Uiwang 16082, Republic of Korea
}

\author{Seunghyo Noh}
\affiliation{ 
  Materials Research \& Engineering Center, CTO Division, Hyundai Motor Company, Uiwang 16082, Republic of Korea
}

\author{Young Min Rhee}
\affiliation{Department of Chemistry, KAIST, Daejeon, 34141, Republic of Korea}

\author{Kyungmin Kim}
\affiliation{Department of Chemistry, KAIST, Daejeon, 34141, Republic of Korea}

\date{\today}

\begin{abstract}

The electron pair approximation offers a resource efficient variational quantum eigensolver (VQE) approach for quantum chemistry
simulations on quantum computers. With the number of entangling gates scaling quadratically with system size and a constant energy measurement overhead, the orbital optimized unitary pair coupled cluster double (oo-upCCD) ansatz strikes a balance between accuracy and efficiency on today's quantum computers. However, the electron pair
approximation makes the method incapable of producing quantitatively accurate energy predictions. In order to improve the accuracy without increasing the circuit depth, we explore the idea of reduced density matrix (RDM) based 
second order perturbation theory (PT2) as an energetic correction to electron pair approximation. The new approach takes into account of the broken-pair energy contribution that is missing in pair-correlated electron simulations, while maintaining the computational
advantages of oo-upCCD ansatz. In dissociations of N$_2$, Li$_2$O, and chemical reactions such as the unimolecular decomposition of CH$_2$OH$^+$ and the \snTwo reaction of CH$_3$I $+$ Br$^-$, the method significantly improves the accuracy of energy prediction. 
On two generations of the IonQ's trapped ion quantum computers, Aria and Forte, we find that unlike the VQE energy, the PT2 energy correction is highly noise-resilient. By applying a simple error mitigation approach based on post-selection solely on the VQE energies, the 
predicted VQE-PT2 energy differences between reactants, transition state, and products are in excellent agreement with noise-free simulators. 

\end{abstract}

\maketitle

\section{Introduction}
Quantum computers have the potential to revolutionize a number of fields, including physical simulation,\cite{Aspuru-Guzik19_10856} optimization,\cite{Gutman14_14114028} and machine learning.\cite{Fiorentini19_043001} Of particular interest is to solve the electronic
structure problem of molecules and materials using quantum computers. It is well known
that classically computing exact solutions to the electronic structure problem requires computational
resources that grow exponentially with system size. To mitigate this problem, the current state of the art is to use polynomial scaling approximate methods to predict electronic properties. One example is the ``gold standard'' of electronic structure theory, projected coupled-cluster theory\cite{Szabo-Ostland}, namely CCSD(T). 

In contrast, quantum algorithms unlock the potential to solve the \emph{exact} electronic structure problem to arbitrary accuracy in polynomial space and time. Unfortunately, known algorithms with robust theoretical guarantees (such as phase estimation) remain highly sensitive to noise, and therefore require the use of fault-tolerant quantum computers with error corrections. Fault-tolerant quantum devices are likely many years away, so in the interim much effort has been made to find algorithms that are noise-robust but still provide accurate solutions to electronic 
properties. The ultimate goal is to enable reliable predictions of energetics, geometries, binding affinities, response properties, and more. The reason for this is that access to robust and dependable predictions promises to 
greatly aid the development of functional materials,\cite{Chan20_12685} next-generation battery,\cite{Garcia21_134115} drug
discovery,\cite{Holzmann22_00551} and novel catalysts.\cite{Troyer21_033055} Therefore, the potential advantage of quantum computers in this field is significant. 

In recent years, significant progress has been made in quantum technologies, marked by advancements in both algorithmic methods and hardware capabilities. Presently, a variety of cloud-based quantum computing platforms are available, offering access to quantum processors with diverse architectures. These platforms, provided by several commercial cloud vendors, offer a variety of computational capacities, allowing the user to choose between systems with varying qubit connectivities, gate operation times, fidelity measures, and qubit counts. This expansion in hardware has been paralleled by a correspondingly rapid development of algorithms for quantum chemistry simulations, which are increasingly being implemented and tested on these emerging quantum computational systems.

Since the original proposal of using 
the quantum phase estimation (QPE)\cite{Head-Gordon05_1704, white10_106} approach to estimate ground state energies of molecules, 
researchers have developed even more efficient implementations, through advances such as qubitization\cite{Chuang19_163,Neven18_041015} and to leverage factorization methods such as  tensor hypercontraction\cite{Babbush21_030305}. Using these algorithms, researchers have provided numerous estimates of the
resources\cite{Troyer21_033055, Troyer17_7555, Holzmann22_7001}, such as compute time and number of qubits, required
to solve some of the hardest chemical problems exactly with a fault-tolerant quantum computer.

Although resource estimation for fault-tolerant quantum computing is crucial to the development of the field, broadly accessible (and useful) fault-tolerant quantum computers are likely many years away. Currently, quantum devices operate within the era of noisy 
intermediate scale quantum (NISQ)\cite{Preskill18_79, Aspuru-Guzik22_015004} computing. In the NISQ era, quantum computers face limitations in the number of utilizable qubits and the fidelity of quantum gates. For example, 
the Aria quantum computer developed by IonQ has 25 qubits with a two-qubit
gate fidelity of 99.4\%. Despite the high fidelity, the chance of error remains nonzero and this limits the number of operations that can be performed reliably on QPUs. This in turn limits the design and application of quantum algorithms. Developing algorithms that are noise-resilient on NISQ systems 
has drawn significant attention, and remains of utmost importance in the near-term. For chemistry, one NISQ algorithm that is particularly promising
is the variational quantum eigensolver (VQE)\cite{Brien14_5213, Martinis16_031007, Gambetta17_23879,Google20_1084,Kim20_33, Nam21_04151} and its variants. VQE utilizes parametrized quantum circuits, known as ansatze, to estimate electronic wave functions. It employs the variational principle to optimize these parameters so as to locate the energy minimum. This minimum represents the optimal approximation of the exact wave function, within the limits of the parametrized circuit. VQE offers users the choice of an ansatz to strike a balance between accuracy and fidelity. Simpler ansatze lead to more efficient, shallower circuits, which are amenable to NISQ systems. However, they may lack the expressiveness to capture accurate chemical behaviors. Conversely, complex ansatze can improve accuracy, but their effectiveness is often diminished by noise as a consequence of compounding numerous qubit operations. Therefore, given a certain 
number of qubits and gate fidelities, users must carefully design VQE circuits
to strike a balance between accuracy and fidelity. 
The potential quantum advantage of VQE 
is that it allows the efficient utilization of ansatze that are inefficient to compute classically, 
such as the unitary coupled cluster (UCC) ansatz,\cite{Mayhall20_1,Mayhall19_3007, Brien14_5213, Martinis16_031007, Pooser19_99}. The UCC class of ansatze is known to offer more accurate 
energy predictions to its classical (projective) counterpart, especially for strongly correlated
systems. However, as with the classical case, care must be given to the design and selection of the UCC-based ansatz depending on the chemical case at hand.

We recently developed a resource-efficient VQE algorithm\cite{Zhao22_60} for quantum chemistry simulations. This algorithm takes advantage of the unitary pair coupled cluster doubles
(upCCD)\cite{Tavernelli20_124107, Rubin22_10799, Neuscamman16_5841} ansatz. In the upCCD ansatz, all the electrons are paired, which allows each electron pair to be treated as an effective (hard-core) bosonic particle. Computationally, this allows one to map spatial orbitals to individual qubits. In doing so, the number of qubits required are reduced by half 
compared to the conventional Jordan-Wigner mapping of spin orbitals to qubits. In previous work, we demonstrated that the upCCD ansatz 
yields highly efficient circuits with a constant overhead for energy measurements. We
further showed that the orbital optimization (oo) effects, which are crucial for 
modeling bond breaking scenarios, could be applied to the upCCD circuit for no additional quantum cost, in the sense that there is no increase in either the circuit depth or the number of measurements beyond the additional macro-iterations for the orbital optimization loop. Preliminary results 
from using the oo-upCCD ansatz on IonQ's trapped-ion quantum computer show that the 
predicted relative energies are in excellent agreements with a noiseless quantum simulator. 

Despite this progress, it is worth noting that the oo-upCCD ansatz is not able to produce quantitative
levels of accuracy, except for two-electron problems. The main reason is that it 
restricts the electrons to pairs, and the (generally significant) correlation energy contribution from 
broken-pair excitations is ignored. How one could account for the broken-pairs for 
oo-upCCD thus becomes a pressing issue. One obvious route is to implement these excitations
using quantum circuits, but doing so one would immediately sacrifice two major advantages of the 
oo-upCCD ansatz: 1) the ability to only map spatial orbitals to qubits, and, 2) the constant 
energy measurement overhead. In addition, it would also drastically increase the circuit
depth, even with the most optimal circuit compilation techniques.\cite{Goings23_00667} 

In this study, we pursue an alternative approach by first computing the reduced density matrices (RDMs) of the oo-upCCD circuits. From these RDMs, we leverage a second-order perturbation theory (PT2) framework to calculate the energetic contributions from broken-pair excitations. This method does not add to the circuit depth and maintains a constant energy measurement overhead. We apply this technique to model bond dissociation and chemical reactions using quantum simulators and IonQ's trapped-ion quantum computers. Our objectives are twofold: to determine the accuracy achievable using oo-upCCD with a perturbative correction, as well as assess the performance of the perturbative algorithm on current quantum computing hardware. 

The paper is structured as follows, we begin by introducing the oo-upCCD ansatz and
its circuits. Then we dive into the second order perturbation theory and discuss
the zeroth order Hamiltonian, the wave function correction, and derive the working 
equation for wave function and energy corrections.  
Having laid out the general formalism, we shift our attention to some of the numerical 
issues we face in the algorithms and some proposed solutions to it. 
Results are presented on quantum simulators and IonQ's Aria and Forte quantum computers for potential energy surface predictions of N$_2$,  
Li$_2$O, and two chemical reactions. We conclude with 
a summary of our findings and comments on future directions.


\section{Theory}
\subsection{The oo-upCCD Ansatz}

The unitary pair coupled-cluster double (upCCD) ansatz is 
\begin{equation}
    \label{eqn:pccd_ansatz}
    \left|\Psi_{\mathrm{u}\mathrm{pCCD}}\right>=e^{T-T^\dagger}\left|\mathrm{HF}\right>
\end{equation}
in which $T$ is the pair-double cluster operator, defined as
\begin{equation}
    T=\sum_{ia}{t_i^a a_{a\alpha}^\dagger a_{a\beta}^\dagger a_{i\beta}a_{i\alpha}}
\end{equation}
in which $i$ and $a$ are indices for occupied and unoccupied orbitals in the HF state. 
$a^\dagger_{p\alpha}$ ($a^\dagger_{p\beta}$) and $a_{p\alpha}$ ($a_{p\beta}$) are the 
fermionic creation and annihilation operators in the $p$-th spin-up ($\alpha$) or spin-down ($\beta$) orbital. 

\begin{figure}[h!]
\centering
\includegraphics[width=8.5cm,angle=0,scale=1.0]{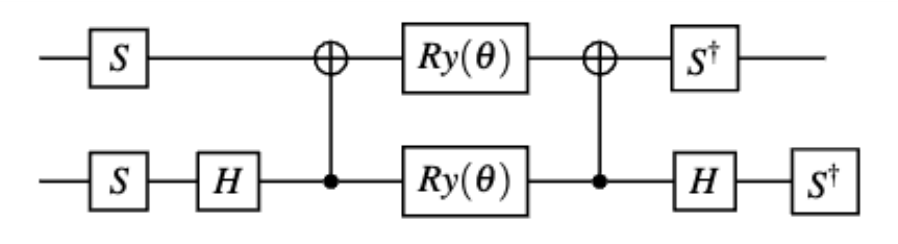}
\caption{
  A quantum circuit that implements the Givens rotation.  
        }
\label{fig:givens_rotation_cirq}
\end{figure}

The exponentiation of the pair-excitation operator can be efficiently implemented with 
the following circuit, given in Figure \ref{fig:givens_rotation_cirq}, 

Once the circuit is defined, one needs to measure the energy expectation value $\left<\Psi_{\mathrm{upCCD}}|H|\Psi_{\mathrm{upCCD}}\right>$ for the second-quantized Hamiltonian 
$H$. In the full electronic Hamiltonian, there are $O(N^4)$ terms in $H$, for which $N$ is the number of qubits. However, due to pair symmetry, a majority 
of them do not contribute to the energy in the upCCD approximation. After eliminating these 
pair-breaking terms, one finds that only three measurements are needed (in the $X$, $Y$, and $Z$ basis, respectively) to compute
the energy. Moreover, the number of basis measurements are independent of problem size, and three circuits are all that is needed for \emph{any} upCCD calculation.

One potential concern is that the upCCD ansatz defined in Equation \ref{eqn:pccd_ansatz} is not invariant to the choice of underlying orbitals. 
Previous studies \cite{Bultinck14_853,Scuseria15_214116,Neuscamman16_5841,Tavernelli20_124107} on similar wave functions have found that it is necessary to optimize the orbitals along with the cluster
amplitudes, especially for strongly correlated systems. The orbital optimized upCCD ansatz is
\begin{equation}
    \label{eqn:oo-upccd}
    \left|\Psi_{\mathrm{oo}-\mathrm{upCCD}}\right>=e^Ke^{T-T^\dagger}\left|\mathrm{HF}\right>
\end{equation}
in which there are two different sets of parameters: 1) circuit parameters in the cluster operator $T$, and, 2) orbital rotation parameters in the the orbital rotation operator $K$, which is defined as 
\begin{equation}
    \label{eqn:k_op}
    K=\sum_{p>q}{\sum_{\sigma}{K_{pq}(a^\dagger_{p\sigma}a_{q\sigma}-a^\dagger_{q\sigma}a_{p\sigma})}}
\end{equation}
Here $K$ is an anti-Hermitian matrix and $\sigma$ indexes the spin. 

\subsection{The VQE Perturbation Theory}
As mentioned previously, the cluster operator in upCCD does not contain any broken-pair excitations. By neglecting these terms, we observe a quantitative deviation from the exact energy. To remedy this, we would like to recover the energetic corrections from these terms. To account for the broken-pair contributions we use perturbation theory as follows. We first write
the molecular Hamiltonian as 
\begin{equation}
    H=H_0+\lambda V
\end{equation}
in which $H_0$ is a simple non-interacting Hamiltonian and $V$ accounts for the electron interaction. $\lambda$
is the scale of the interaction.

We then write the energy and wave function in terms of $\lambda$
\begin{equation}
    \begin{split}
        &\left|\Psi\right>=\left|\Psi^{(0)}\right>+\lambda\left|\Psi^{(1)}\right>+\lambda^2\left|\Psi^{(2)}\right>+\cdots \\
        &E=E^{(0)}+\lambda E^{(1)}+\lambda^2E^{(2)}+\cdots. \\
    \end{split}
\end{equation}

Plugging the wave function and energy into the time-independent Schrodinger's equation, and selecting terms up to second order in $\lambda$, we obtain
\begin{equation}
\label{eqn:pt2_diff_orders}
    \begin{split}
        &H_0\left|\Psi^{(0)}\right>=E^{(0)}\left|\Psi^{(0)}\right> \\
        &H_0\left|\Psi^{(1)}\right>+V\left|\Psi^{(0)}\right>=E^{(0)}\left|\Psi^{(1)}\right>+E^{(1)}\left|\Psi^{(0)}\right> \\
        &H_0\left|\Psi^{(2)}\right>+V\left|\Psi^{(1)}\right>=E^{(0)}\left|\Psi^{(2)}\right>+E^{(1)}\left|\Psi^{(1)}\right>+E^{(2)}\left|\Psi^{(0)}\right> \\
    \end{split}
\end{equation}

Next, we constrain $\left|\Psi^{(1)}\right>$ such that it is orthogonal to $\left|\Psi^{(0)}\right>$
\begin{equation}
    \left<\Psi^{(1)}|\Psi^{(0)}\right>=0. 
\end{equation}

After left projecting Equation \ref{eqn:pt2_diff_orders} by $\left<\Psi^{(0)}\right|$, we obtain
\begin{equation}
    \begin{split}
        &\left<\Psi^{(0)}|H_0|\Psi^{(0)}\right>=E^{(0)} \\
        &\left<\Psi^{(0)}|V|\Psi^{(0)}\right>=E^{(1)} \\
        &\left<\Psi^{(0)}|V|\Psi^{(1)}\right>=E^{(2)} \\
    \end{split}
\end{equation}

For our reference (zeroth-order) wave function, $\left|\Psi^{(0)}\right>$, we can use the oo-pUCCD wave function obtained by oo-VQE:

\begin{equation}
\left|\Psi^{(0)}\right>=
\left|\Psi_{\mathrm{oo-upCCD}}\right>
\end{equation}

We then partition the second-quantized molecular Hamiltonian as
\begin{align}
    H = \sum_p{\epsilon_pa^\dagger_pa_p} + \sum_{pq}h_{pq} a_p^\dagger a_q + \sum_{pqrs} g_{pqrs} a_p^\dagger a_q a_r^\dagger a_s
\end{align}

and we define the zeroth order Hamiltonian as
\begin{equation}
    H_0=\sum_p{\epsilon_pa^\dagger_pa_p}
\end{equation}
in which where $p$ runs over all spin orbitals and  $\epsilon_p$ are the orbital energies 
computed by plugging the one body reduced density matrix (1-RDM) of oo-upCCD into the HF
orbital energy expression. 
Similarly, we define the perturbation potential $V$ as
\begin{equation}
\label{eqn:perturb-potential}
    V=\sum_{pq}h_{pq} a_p^\dagger a_q + \sum_{pqrs} g_{pqrs} a_p^\dagger a_q a_r^\dagger a_s.
\end{equation}

The VQE energy on its own contains both the zeroth and first order energy corrections
\begin{equation}
    E_{\mathrm{VQE}}=E^{(0)}+E^{(1)}.
\end{equation}
The leading energy correction beyond the VQE energy enters in the second order
\begin{equation}
    \begin{split}
        E_{\mathrm{VQE}}=E^{(0)}+E^{(1)} &= \left<\Psi^{(0)}|H_0|\Psi^{(0)}\right>+ \left<\Psi^{(0)}|V|\Psi^{(0)}\right>\\
        &=\left<\Psi^{(0)}|H_0+V|\Psi^{(0)}\right> \\
        &=\left<\Psi^{(0)}|H|\Psi^{(0)}\right>.
    \end{split}
\end{equation}

Let's now write the first-order wave function correction as 
\begin{equation}
\label{eqn:1st-wfn-perturb}
    \left|\Psi^{(1)}\right>=\sum_{pqrs}^{non-pair}{t_{pqrs}a^\dagger_pa_qa^\dagger_ra_s\left|\Psi^{(0)}\right>}=\sum_{P}^{non-pair}{t_{P}\left|\Psi_P\right>}
\end{equation}
in which $a^\dagger_pa_qa^\dagger_ra_s$ breaks electron pairs. 

In order for $a^\dagger_pa_qa^\dagger_ra_s$ to break electron pairs, we must consider these three scenarios
\begin{equation}
    \begin{split}
        &a^\dagger_{p\alpha}a_{q\alpha}a^\dagger_{r\alpha}a_{s\alpha} (r\neq s, p\neq q, r\neq q, s\neq p) (p<r,q<s) \\
        &a^\dagger_{p\beta}a_{q\beta}a^\dagger_{r\beta}a_{s\beta} (r\neq s, p\neq q, r\neq q, s\neq p) (p<r,q<s)\\
        &a^\dagger_{p\alpha}a_{q\alpha}a^\dagger_{r\beta}a_{s\beta} (r\neq s, p\neq q, p\neq r || q\neq s) \\
    \end{split}
\end{equation}
in which the first parenthesis enforces broken-pairing, and the second parenthesis ensures no double counting. 

The second-order energy correction becomes
\begin{equation}
\label{eqn:ept2}
    E^{(2)}=\sum_P^{non-pair}{t_P\left<\Psi^{(0)}|V|\Psi_P\right>}
\end{equation}
in which we need to compute $t_P$. 

Left projecting Equation \ref{eqn:pt2_diff_orders} by $\left|\Psi_Q\right>$, we obtain 
\begin{equation}
    \sum_P{t_P\left<\Psi_Q|H_0|\Psi_P\right>}+\left<\Psi_Q|V|\Psi^{(0)}\right>=E^{(0)}\sum_{P}{t_P\left<\Psi_Q|\Psi_P\right>}
\end{equation}
where both $P$ and $Q$ denote non-pair excitations.
Rearranging the above equation gives:
\begin{equation}
    \begin{split}
    &\sum_P t_P\left(
    {
    \left<\Psi_Q|H_0|\Psi_P\right>}
    -E^{(0)}{\left<\Psi_Q|\Psi_P\right>}\right)
    =-\left<\Psi_Q|V|\Psi^{(0)}\right> \\
    &\sum_P  G_{PQ} t_P= - Y_Q 
    \end{split}
\end{equation}
where we define
\begin{equation}
    \begin{split}
        &G_{PQ}=\left<\Psi_Q|H_0|\Psi_P\right>-E^{(0)}\left<\Psi_Q|\Psi_P\right> \\
        &Y_P=\left<\Psi_P|V|\Psi^{(0)}\right>
    \end{split}
\end{equation}
so we can then solve $\vec{t}$ by 
\begin{equation}
    \vec{t}=-\textbf{G}^{-1}\vec{Y}.
    \label{eqn:pt2_amplitude}
\end{equation}

We then plug $\vec{t}$ into Equation \ref{eqn:ept2} to compute $E^{(2)}$. Finally, the total energy is
\begin{equation}
    E=E_{\mathrm{VQE}}+E^{(2)}
\end{equation}

Now we need to compute the $\textbf{G}$ matrix and the $\vec{Y}$ vector. Elements of the $\textbf{G}$ matrix take the form
\begin{equation}
    \begin{split}
    G_{PQ}&=\sum_k{\epsilon_k\left<\Psi^{(0)}|a^\dagger_ua_va^\dagger_xa_ya^\dagger_ka_ka^\dagger_pa_qa^\dagger_ra_s|\Psi^{(0)}\right>}\\
    &-E^{(0)}\left<\Psi^{(0)}|a^\dagger_ua_va^\dagger_xa_ya^\dagger_pa_qa^\dagger_ra_s|\Psi^{(0)}\right>
    \end{split}
\end{equation}

We approximate $\textbf{G}$ with its diagonal elements 
\begin{equation}
    \begin{split}
    G_{PP}&=\sum_k{\epsilon_k\left<\Psi^{(0)}|a^\dagger_sa_ra^\dagger_qa_pa^\dagger_ka_ka^\dagger_pa_qa^\dagger_ra_s|\Psi^{(0)}\right>} \\
    &-E^{(0)}\left<\Psi^{(0)}|a^\dagger_sa_ra^\dagger_qa_pa^\dagger_pa_qa^\dagger_ra_s|\Psi^{(0)}\right>\\
    &=\sum_k{\epsilon_k\Gamma_{sqkpr}^{rpkqs}}-E^{(0)}\Gamma_{sqpr}^{rpqs}
    \end{split}
\end{equation}
in which we have defined the 5-particle reduced density matrix (5-RDM). 

We also need to compute the $\vec{Y}$ vector as
\begin{equation}
    \begin{split}
        Y_P&=\sum_{pq}{h_{pq}\left<\Psi^{(0)}|a^\dagger_ua_va^\dagger_xa_ya^\dagger_pa_q|\Psi^{(0)}\right>} \\
        &+\sum_{pqrs}{g_{pqrs}\left<\Psi^{(0)}|a^\dagger_ua_va^\dagger_xa_ya^\dagger_pa_qa^\dagger_ra_s|\Psi^{(0)}\right>} \\
        &=\sum_{pq}{h_{pq}\Gamma_{uxp}^{vyq}}+\sum_{pqrs}{g_{pqrs}\Gamma_{uxpr}^{vyqs}} \\
    \end{split}
    \label{eqn:pt2_y_vec}
\end{equation}
in which we need 3- and 4-RDMs. $h$ and $g$ are the one- and two-electron integrals, respectively. 

The most computationally-intensive step of the VQE-PT2 approach is in the 
computation of the 4-RDMs from Equation \ref{eqn:pt2_y_vec}, which contains eight different indices. However, 
although the term $a^\dagger_pa_qa^\dagger_ra_s\left|\Psi^{(0)}\right>$ in the first-order 
wave function correction contains broken-pair configurations, all electrons should still be paired
in $a^\dagger_ua_va^\dagger_xa_ya^\dagger_pa_qa^\dagger_ra_s\left|\Psi^{(0)}\right>$. Otherwise
the overlap with $\left|\Psi^{(0)}\right>$ becomes zero. Therefore, only four indices out of the
original eight are unique, and the operator $a^\dagger_ua_va^\dagger_xa_ya^\dagger_pa_qa^\dagger_ra_s$ must
be written in terms of 
\begin{itemize}
    \item number operators $n_p=a^\dagger_pa_p\rightarrow\frac{1}{2}(1-Z_p)$
    \item double creation operators $d_p^\dagger=a_{p\alpha}^\dagger a_{p\beta}^\dagger\rightarrow\frac{1}{2}(X_p-iY_p)$
    \item double annihilation operators $d_p=a_{p\beta} a_{p\alpha}\rightarrow\frac{1}{2}(X_p+iY_p)$
\end{itemize}
in which the expression at the right of the arrow is the Jordan-Wigner transformed formalism 
of these operators. 

Therefore, for 4-RDMs, one only needs to measure a subset of rank-4 Pauli strings that includes:
1) Pauli strings with four Pauli-$Z$ matrices, 2) Pauli strings with two Pauli-$Z$ matrices and two 
Pauli-$X$ matrices, 3) Pauli strings with two Pauli-$Z$ matrices and two 
Pauli-$Y$ matrices, 4) Pauli strings with two Pauli-$X$ matrices and two 
Pauli-$Y$ matrices. The rest of the Pauli strings do not contribute as a result of symmetry violation.
We then group this Pauli strings and find terms that are qubitwise commutative\cite{Chong19_13623} with 
each other, and measure them altogether. A detailed description of how to
construct the 4-RDMs efficiently can be found in the Supplementary Information. 

\subsection{Regularization}
The perturbation theory described in previous sections may run into numerical instability
issues without regularization. The numerical instability arises from two reasons. First, 
if a broken-pair excitation in the wave function corrections excites an electron from 
(to) a nearly empty (occupied) orbital, then both $G$ and $Y$ would be small and their 
division in Equation \ref{eqn:pt2_amplitude} would be unstable. Second, since oo-upCCD includes orbital
optimization effects, in bond breaking scenarios, the HOMO-LUMO gap closes as one 
stretches the bond, which also reduces the magnitude of $G$ and leads to numerical instability.

In order to tackle this problem, we take the same approach of the orbital optimized
M{\o}ller-Plesset second order perturbation theory (ooMP2)\cite{Head-Gordon18_5203, Head-Gordon21_12084} and add a regularization to the amplitude, which becomes 
\begin{equation}
    t_p=-\frac{Y_p}{G_{pp}}(1-e^{-\sigma G_{pp}})
\end{equation}
in which we introduce an empirical parameter $\sigma$ that controls the strength of the 
regularization. 

As one could see, if $G=0$, then the regularized $t_p$ also becomes zero and it no longer 
contributes to the wave function. In the regularized ooMP2, $\sigma$ is obtained by fitting
to a known database, and the optimal $\sigma$ is found to be 1.5. In our case, we use a much larger
$\sigma$, which is chosen to be 28. Such a large $\sigma$ allows us to completely remove the 
numerical instability that arises when $G$ is close to zero, while it only damps little 
correlation energy away when $G$ is far away from zero. In the future,
a more appropriate scheme for choosing $\sigma$ may be to fit to a known dataset. However, optimal selection of $\sigma$ is beyond the scope of the present study.

\section{Computational Details}
The oo-upCCD circuits are implemented in the  Qiskit software platform,\cite{Qiskit} and all the simulator results are obtained from the statevector simulator
provided by Qiskit. The full configuration
interaction (FCI) and the complete active space configuration interaction (CASCI) results were obtained from PySCF\cite{pyscf} package. The Li, O, C, and N $1s$ orbitals are frozen at the restricted Hartree-Fock (RHF) level. For Br and I atoms, all the non-valence orbitals are frozen at the RHF level. 
The oo-upCCD circuit parameter optimization is performed with 
the L-BFGS optimization technique. All the calculations are performed using the minimal STO-3G basis. Error bars on the potential energy surface plots 
represent $\pm$ one standard error as a result of finite sampling. The molecular geometries 
along the reaction path for CH$_2$OH$^+$ decomposition were taken from intrinsic reaction coordinate calculation\cite{Kim98_5363}. The molecular structures along the minimum energy path of \snTwo reaction of CH$_3$I $+$ Br$^-$ were obtained using the nudged elastic band method\cite{Jonsson00_9901, Jonsson00_9978, Henkelman08_134106}. 

The experimental demonstration was performed on the Aria and Forte quantum processing units (QPUs) developed by IonQ.  
Both QPUs utilize trapped Ytterbium ions where two states in the ground
hyperfine manifold are used as qubit states. These states are manipulated by illuminating individual ions with
pulses of 355 nm light that drive Raman transitions between the ground states defining the qubit. By configuring these pulses, arbitrary single qubit gates and M{\o}lmer-S{\o}renson type two-qubit gates can both be realized. 
The Forte system\cite{gamble23_05071} contains larger qubit registers and improved gate fidelities than Aria due to the acoustic-optic deflectors that allow independent alignment of each laser beam to each
ion. This technique leads to smaller beam alignment error across the 
chain of trapped ions. 

\section{Results}

\subsection{N$_2$}
\begin{figure}[h!]
\centering
\includegraphics[width=8.5cm,angle=0,scale=1.0]{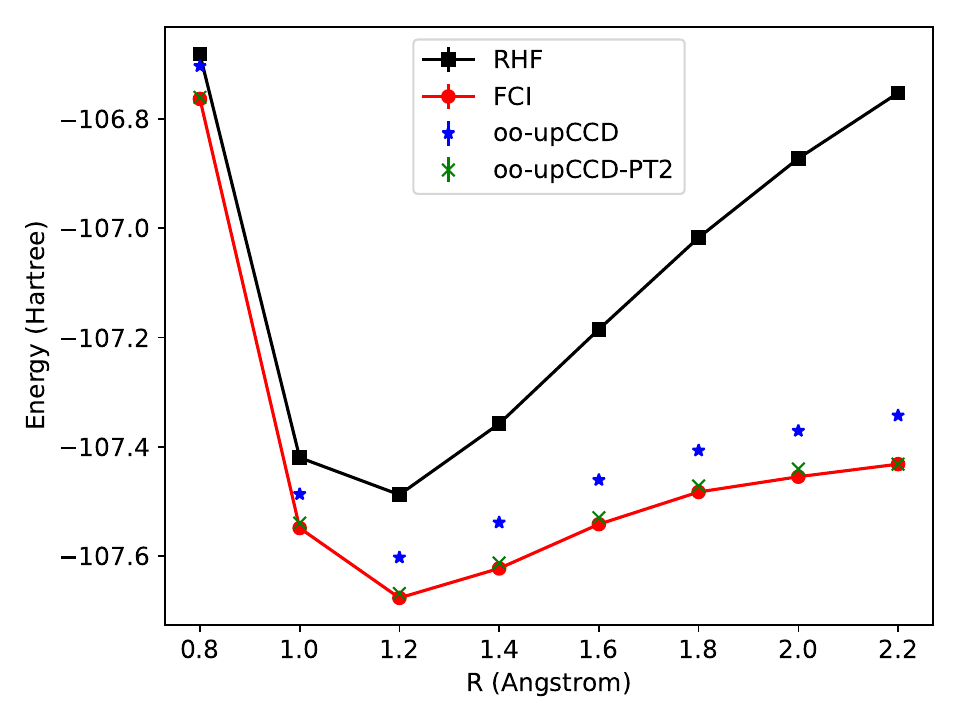}
\caption{
  Potential energy surface for the dissociation of the N$_2$ molecule in the STO-3G basis computed with RHF, oo-upCCD, oo-upCCD-PT2, and FCI. VQE results are obtained from a noise-free simulator. 
        }
\label{fig:n2_simulator}
\end{figure}
We begin our numerical results with the dissociation of the N$_2$ triple bond. In Figure \ref{fig:n2_simulator}, we compare the energy predicted by RHF,  
oo-upCCD, oo-upCCD-PT2, and FCI. We observe that the oo-upCCD energy provides a significant improvement over RHF, but it
remains far from the FCI energy. This is due to the missing of 
broken-pair excitation contributions in the oo-upCCD wave function. After
applying the perturbation correction, we find that it is now much closer to FCI. The nonparallelity error (NPE) defined as the difference between the largest and smallest error with respect to FCI along the potential surface, is reduced from 52mH to 14mH, which demonstrates the effectiveness of the perturbation correction. 

\subsection{Li$_2$O}
\begin{figure}[h!]
\centering
\includegraphics[width=8.5cm,angle=0,scale=1.0]{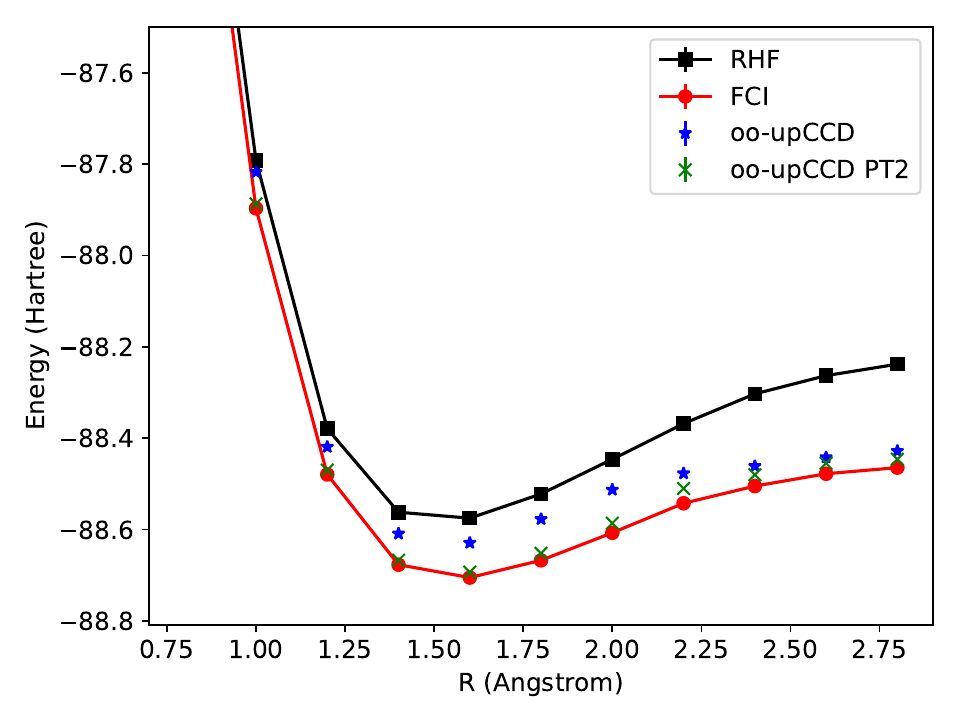}
\caption{
  Potential energy surface for the symmetric dissociation of the Li$_2$O molecule in the STO-3G basis computed with RHF, oo-upCCD, oo-upCCD-PT2, and FCI. VQE results are obtained from a noise-free simulator. 
        }
\label{fig:li2o_simulator}
\end{figure}
Our second example is the symmetric dissociation of the Li$_2$O molecule. Li$_2$O is one of the secondary reaction products in lithium-air batteries, 
which is a potential candidate for next-generation lithium battery due to its high energy density. The results on an ideal simulator are shown in Figure \ref{fig:li2o_simulator}, 
comparing RHF, FCI, oo-upCCD, and oo-upCCD-PT2. As one could see, similar to the N$_2$ 
dissociation, the oo-upCCD-PT2 significantly improves the accuracy of oo-upCCD, and reduces
the NPE v.s. FCI from 69mH to 24mH. This is particularly true in the equilibrium
geometry. When bonds are stretched, oo-upCCD-PT2 still exhibits some noticeable amount of errors compared to
FCI. These remaining errors are due to the limitation of the chosen form of the second order perturbation theory, such as 1) higher order terms are needed 2) one needs to use a different
zeroth-order Hamiltonian than the simple one-body one we use. 3) the diagonal approximation of the 
$G$ matrix. 

\subsection{Chemical Reactions}
\subsubsection{Results on Simulator}
\begin{figure}
\centering
\includegraphics[width=8.5cm,angle=0,scale=1.0]{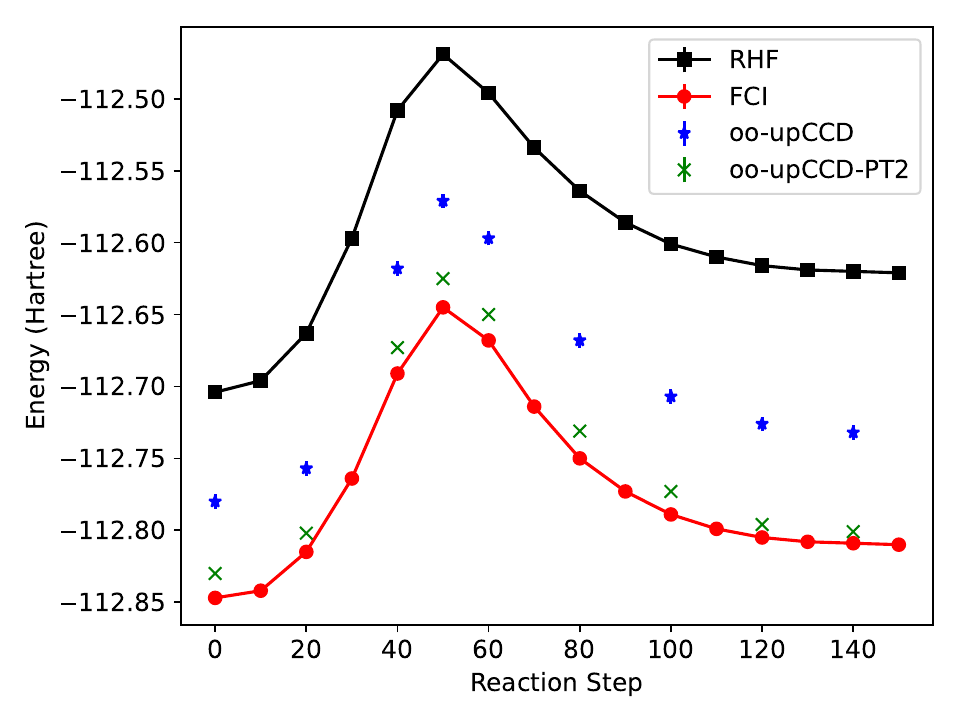}
\caption{
  The simulated CH$_2$OH$^+\rightarrow$ HCO$^+$+H$_2$ reaction energy pathway using RHF, oo-upCCD, oo-upCCD-PT2, and FCI. VQE results are obtained from a noise-free simulator. 
        }
\label{fig:ch2oh_simulator}
\end{figure}
We now move our attention to the chemical decomposition process of the CH$_2$OH$^+\rightarrow$ HCO$^+$+H$_2$ and the \snTwo reaction CH$_3$I $+$ Br$^-\rightarrow$ CH $_3$Br $+$ I$^-$. The simulated energy
profile along the reaction path is shown in Figure \ref{fig:ch2oh_simulator} for the 
CH$_2$OH$^+$ decomposition. After freezing the core orbitals, the remaining eleven spatial
molecular orbitals are mapped to eleven qubits. 
Unsurprisingly, oo-upCCD without perturbative corrections is insufficient
to produce quantitative accuracy for absolute energies, capturing roughly only 50\% of correlation energy. However, due to error cancellation, the
predicted reaction energy barrier (209 mH) is close to FCI (202 mH). The energy 
difference between reactants and products ($\Delta$E) is predicted to be 50 mH by oo-upCCD, which overestimates
the FCI prediction (38 mH) by 24\%. 
Applying the PT2 correction significantly improves the results, and we are able to capture 88\% of the correlation. The predicted reaction energy barrier and $\Delta$E is 205 mH and 39 mH respectively, 
which reduces the original oo-upCCD error by 57\% and 92\%. 

\begin{figure}[h!]
\centering
\includegraphics[width=8.5cm,angle=0,scale=1.0]{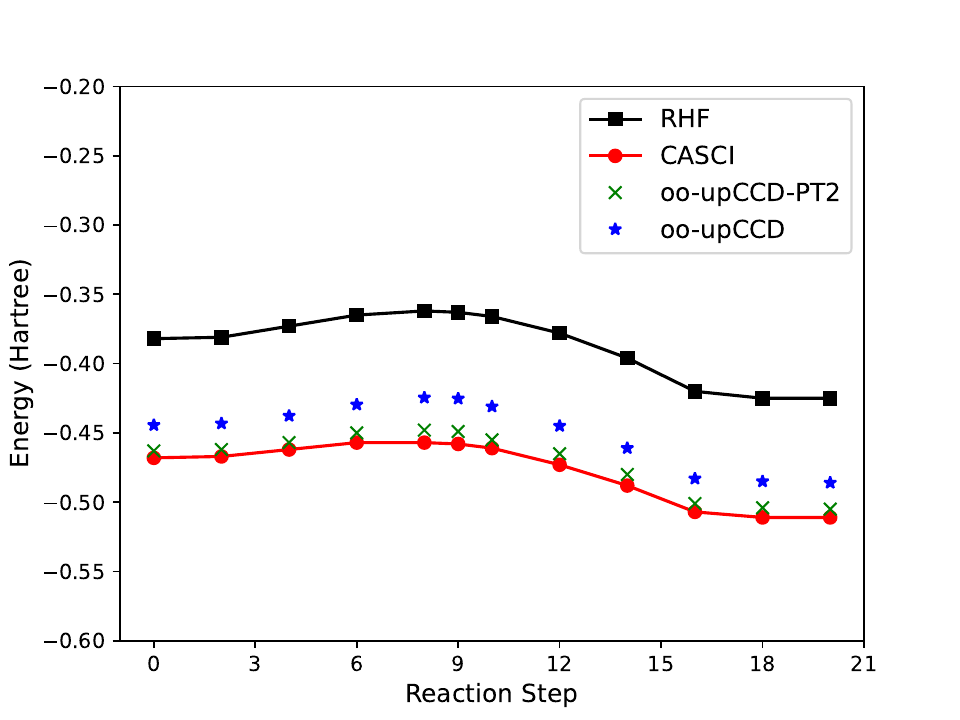}
\caption{
  The simulated CH$_3$I $+$ Br$^-\rightarrow$ CH $_3$Br $+$ I$^-$ \snTwo reaction reaction energy pathway using RHF, oo-upCCD, oo-upCCD-PT2, and FCI. VQE results are obtained from a noise-free simulator.  Energies are offset by 9434 Hartree for legibility.
        }
\label{fig:sn2_simulator}
\end{figure}

The simulated results of the \snTwo reaction of CH$_3$I $+$ Br$^-$ are shown in Figure \ref{fig:sn2_simulator}. 
After freezing the core orbitals, we end up with 15 spatial 
orbitals/qubits to simulate. We now use the CASCI simulations in the same active space
as references. Similar to the CH$_2$OH$^+$ decomposition, the 
perturbation correction brings the oo-upCCD energy much closer to the CASCI predictions. Besides 
absolute energies, relative energies are also improved. The predicted barrier height is 14 mH by oo-upCCD-PT2 v.s. 10 mH by CASCI, which reduces the error of oo-upCCD (19 mH) by 50\%. 

\subsubsection{Results on IonQ Quantum Computer}


\begin{figure}[h!]
\centering
\includegraphics[width=8.5cm,angle=0,scale=1.0]{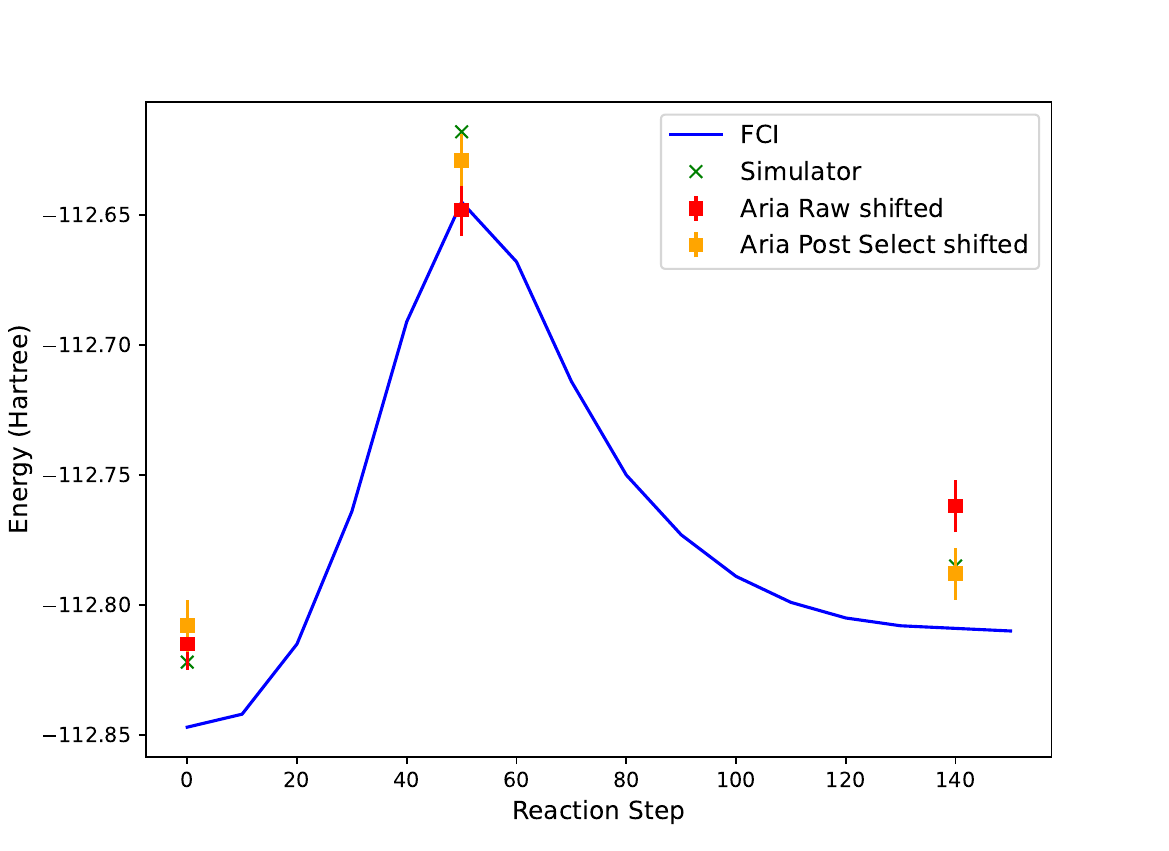}
\caption{
  The simulated CH$_2$OH$^+\rightarrow$ HCO$^+$+H$_2$ reaction energy pathway on the IonQ Aria
  quantum computer. 
        }
\label{fig:ch2oh_aria}
\end{figure}

\begin{figure}[h!]
\centering
\includegraphics[width=8.5cm,angle=0,scale=1.0]{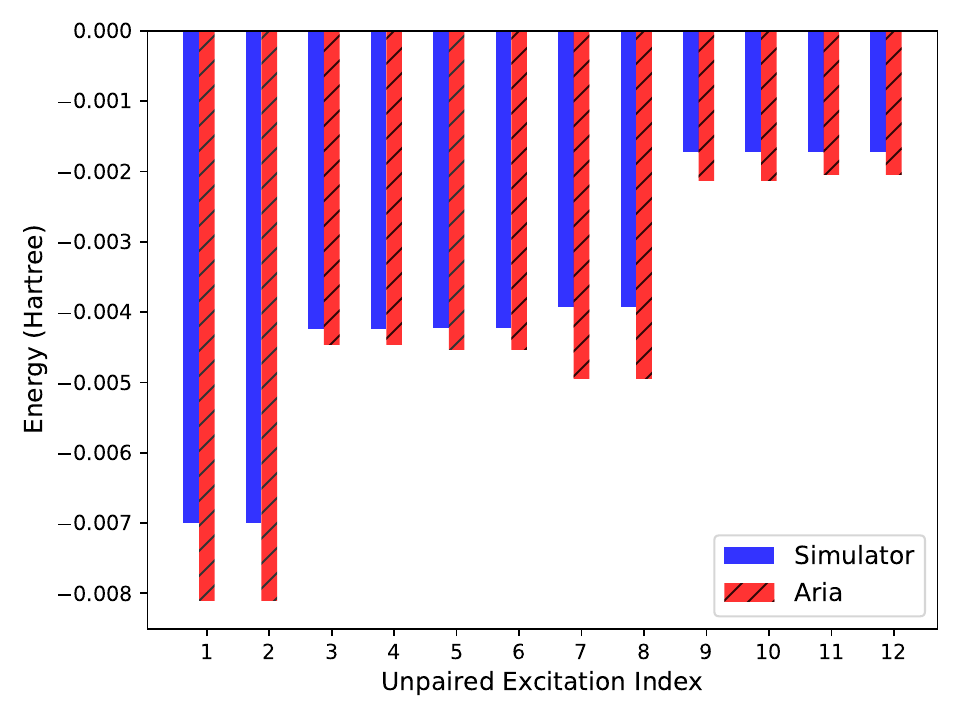}
\caption{
  The energy contribution to E$^{(2)}$ from the top 12 unpaired excitations for 
  reaction step 140 of the CH$_2$OH$^+$ decomposition, comparing the statevector
  simulator and the Aria QPU. 
        }
\label{fig:ch2oh_term_compare}
\end{figure}
Motivated by the success of perturbation theory for oo-upCCD in quantum simulators, we
performed these chemical reaction simulations on two generations of
the IonQ's quantum computers. 
We simulate reaction steps 0, 50, and 140 of the CH$_2$OH$^+$ decomposition process
on the Aria QPU. These three points correspond to the
reactants, transition state, and products, respectively. 
We first perform a circuit pruning process, in which gates whose parameters 
are below a chosen threshold are removed from the circuit. In our experience, the minor energy benefits derived from these small-parameter quantum operations are overshadowed by the introduction of system noise. In our study, we choose the threshold to be 
0.04 radians. The pruned circuits have 10 $CX$ gates and 46 single qubit gates. The experimental 
results are shown in Figure \ref{fig:ch2oh_aria}. As expected, the introduction of hardware noise
yields a systematic, positive bias to the total energy. However, after we shift the 
data points by a constant (364 mH), the experimental results roughly match the prediction of the noiseless simulator. This suggests that the hardware errors are consistent 
across reaction steps and the relative energy is not affected. 

\begin{table}
\caption{
  Energy contributions of oo-upCCD-PT2 for CH$_2$OH$^+$ between the ideal simulator and the Aria QPU. 
  \label{tab:energy_contribution_ch2oh}
}
\begin{tabular}{ c c c c c}
\hline\hline
 System & $E_{\mathrm{VQE}}$ (sim) & $E_{\mathrm{VQE}}$ (Aria) & $E^{(2)}$ (sim) & $E^{(2)}$ (Aria) \\
 \hline
 Reactant & -112.822 & -112.400 & -0.048 & -0.051 \\  
 Transition State & -112.565 & -112.228 & -0.053 & -0.057 \\
 Product & -112.726 & -112.332 & -0.059 & -0.067  \\
 \hline\hline
\end{tabular}
\end{table}

\begin{table}
\caption{
  Energy contributions of oo-upCCD-PT2 for the \snTwo reaction between the ideal simulator and the Forte QPU. 
  \label{tab:energy_contribution_sn2}
}
\begin{tabular}{ c c c c c}
\hline\hline
 System & $E_{\mathrm{VQE}}$ (sim) & $E_{\mathrm{VQE}}$ (Forte) & $E^{(2)}$ (sim) & $E^{(2)}$ (Forte) \\
 \hline
 Reactant         & -9434.463 & -9434.014 & -0.018 & -0.019 \\  
 Transition State & -9434.449 & -9433.976  & -0.023 & -0.023 \\
 Product          & -9434.505 & -9433.967  & -0.019 & -0.019  \\
 \hline\hline
\end{tabular}
\end{table}

The total energy of oo-upCCD-PT2 contains two parts: the oo-upCCD energy ($E_{\mathrm{VQE}}$) and the energy
correction ($E^{(2)}$). Table \ref{tab:energy_contribution_ch2oh} shows the energy contributions from
these two parts comparing the statevector simulator and the Aria QPU, and we find that almost
all the errors in energy are from the 
the oo-upCCD energy term. This could be understood in two ways. First, the magnitude of the oo-upCCD 
energy is much larger than the energy correction, and so with the same error rate, the former
would result in larger absolute errors than the latter. Second, the perturbative energy correction
also benefits from the error cancellation, as the errors in the numerator ($Y_p^2$) and 
the denominator ($G_p$) may cancel with each other. Therefore, the energy correction
appears significantly more resilient to errors than the oo-upCCD energy. In Figure \ref{fig:ch2oh_term_compare} we plot the energy contribution to $E^{(2)}$ from 
unpaired excitations whose $Y_p/G_p^2$ are above 1 mH, comparing the predictions
of the statevector simulator and the Aria QPU. Clearly, the predictions 
from Aria are in excellent agreement with the simulator. 

\begin{figure}[h!]
\centering
\includegraphics[width=8.5cm,angle=0,scale=1.0]{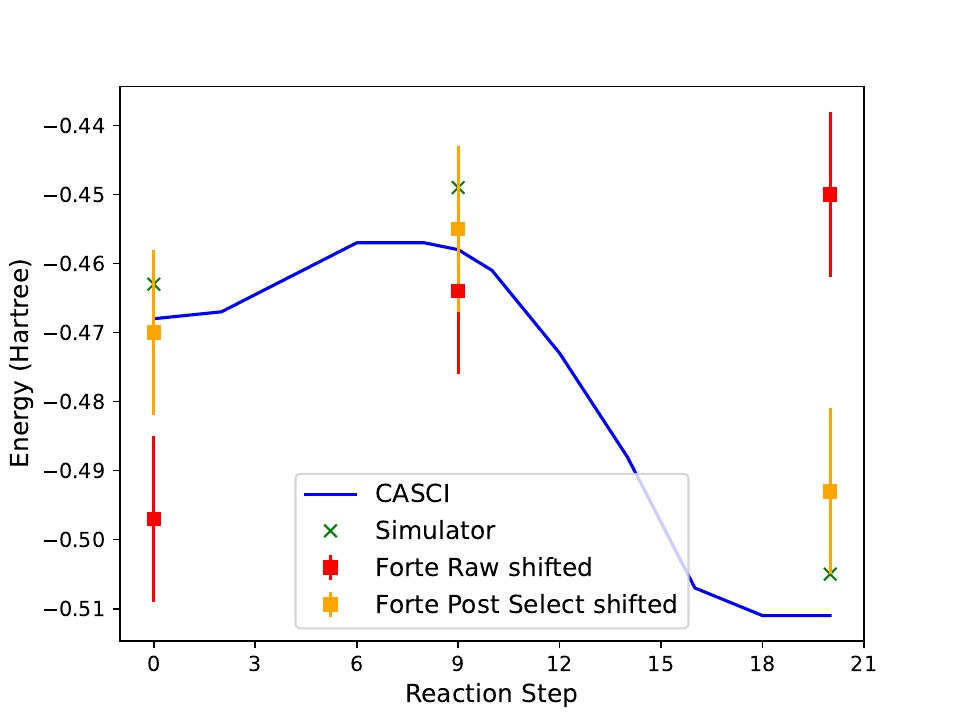}
\caption{
  The simulated CH$_3$I $+$ Br$^-\rightarrow$ CH $_3$Br $+$ I$^-$ \snTwo reaction energy pathway on the IonQ Forte
  quantum computer. Energies are offset 
  by 9434 Hartree for legibility.
        }
\label{fig:sn2_forte}
\end{figure}

In order to improve the absolute energy measurements of oo-upCCD, we take advantage of a simple error mitigation approach based on post selection. For the $Z$-basis measurements, only measurements that preserve particle number symmetry are kept, and the rest are discarded. The results are shown in Figure \ref{fig:ch2oh_aria}. We find that doing so improves the energy measurements by about 200 mH. We then perform the uniform shift for the post-selected energies by 151 mH, and the results better match the results from the ideal simulator. This demonstrates that the post-selection not only improves absolute energies, but also relative energies.

The experimental results of the CH$_3$I $+$ Br$^-$ \snTwo reaction are shown in Figure \ref{fig:sn2_forte} and Table \ref{tab:energy_contribution_sn2}. 
These results are obtained using the IonQ Forte QPU. The \snTwo reaction of CH$_3$I $+$ Br$^-$ is 
particularly challenging for NISQ quantum systems as the energy differences across the 
reaction path are small. The energy difference between the transition state and the 
reactant is about 50 mH, which is a value on the order of the experimental uncertainty. As seen in 
Figure \ref{fig:sn2_forte}, the raw energy demonstrates nonphysical behavior. That is, the product energy is 
even higher than the transition state and the reactant energies. Similar to the CH$_2$OH$^+$ decomposition, 
the error in the total energy is predominantly due to errors from the unperturbed 
VQE energy. Therefore, we apply the same $Z$-basis post selection, and find that the
error mitigated energies now yield the correct behavior, matching well with the 
predictions of the simulator and CASCI.

\section{Conclusion}
A common question regarding current NISQ quantum computers is their utility for quantum chemistry, particularly for electronic structure problems. To address this, accurate simulations on these systems are essential. However, considering the VQE algorithm as an example, many proposed VQE algorithms—based on unitary coupled cluster ansatz or hardware-efficient ansatz,\cite{Gambetta17_23879,Taverneli18_022322,Izmaylov18_6317,Izmaylov20_1055,Parrish21_113010} still produce circuits that are beyond the capabilities of today's 
quantum computers, except for very small chemical systems. The oo-upCCD ansatz is arguably the most practical VQE variant for execution on contemporary quantum computers for chemical systems. Its benefits are threefold: first, it requires only half the qubits compared to many existing VQE ansatze. Second, the scaling of two-qubit gates with system size is quadratic ($O(N^2)$) with respect to system size. Finally, the number of necessary circuits to execute remains constant ($O(1)$), regardless of the system size.

Our goal in this study was to explore ways to enhance the precision of the oo-upCCD method while keeping the complexity of the quantum circuit the same (or increasing it in a shallow, controlled manner). Here, we investigated the use of RDM-based second-order perturbation theory. Our findings indicate that this technique can correct most of the errors caused by the electron-pair assumptions of oo-upCCD, especially during the dissociation of multiple bonds in nitrogen (N$_2$) and lithium oxide (Li$_2$O) molecules, as well as in calculating the energy barriers and differences between reactants and products in chemical reactions. Trials conducted on IonQ's trapped-ion quantum computers demonstrate that the energy corrections predicted by PT2 are robust against noise, which can be attributed to the cancellation of errors within the PT2 energy formula. This characteristic suggests that it can be an effective method for use in the current generation of quantum computers.

Despite its advantages, the oo-upCCD-PT2 method faces challenges similar to those found in traditional perturbation theory, including numerical instability and variational collapse. We intend to tackle these challenges in future research. By continuing to refine our methods, in conjunction with hardware improvements, we aim to develop a practical VQE algorithm for near-term devices. This algorithm is intended to solve complex chemical problems with accuracy equal to or surpassing current classical simulation methods. We anticipate that the outcomes of these experiments will offer critical insights into the utility of near-term QPUs for quantum chemistry.

In summary, we have effectively leveraged RDM-based methods to refine the accuracy of the oo-upCCD method through second-order perturbation theory. We demonstrated this through successful applications on both quantum simulators and actual quantum hardware. Motivated by these initial positive results, in future work, we plan to address some of the issues associated with the perturbation theory. A primary issue we observed is the violation of the variational principle, notably during the dissociation of nitrogen (N$_2$). This limitation may be overcome by implementing measurements within the framework of configuration interaction rather than perturbation theory, potentially leading to an approach similar to the quantum subspace expansion (QSE). In addition to QSE, the recently developed non-orthogonal quantum eigensolver (NOQE)\cite{Baek23_030307} also provides with a useful platform for quantum chemistry simulations beyond VQE, although the Hadamard test used in NOQE will increase the circuit complexity. We plan to explore these and more directions with the goal of landing us on practical quantum advantage for quantum chemistry simulations.

\section{Data Availability}
The data presented in this manuscript are available from the corresponding author upon reasonable request.

\begin{acknowledgments}
We thank the Hyundai Motor Company for funding this research through the Hyundai-IonQ Joint Quantum Computing Research Project. We thank Melanie Hiles for running the 
experiments on QPUs. We thank Dr. Seung Hyun Hong, and Dr. Jongkook Lee at Hyundai Motor Company for enlightening discussions.

\end{acknowledgments}

\section*{Author Contributions}
L. Z, K.S., and W. K. conceived of the project. L.Z., J.G., Q. W., and K. S. designed, implemented, and evaluated the algorithms. Experimental data were collected and analyzed by L. Z. and K. S.; All authors contributed to drafting and editing the manuscript. 

\section*{Competing Interests}
The authors declare no competing interests.

\clearpage

\bibliography{Journal_Short_Name, main}

\end{document}


\pagestyle{fancy}
\fancyhf{}
\cfoot{\thepage}

\title[]{Supplementary Information: Enhancing the Electron Pair Approximation with Measurements on Trapped Ion Quantum Computers}

\author{Luning Zhao}
\email{zhao@ionq.co}
\affiliation{
IonQ Inc, College Park, MD, 20740, USA
}

\author{Qingfeng Wang}
\affiliation{
Chemical Physics Program and Institute for Physical Science and Technology, University of Maryland, College Park, MD, 20742, USA
}

\author{Joshua Goings}
\affiliation{
IonQ Inc, College Park, MD, 20740, USA
}

\author{Kyujin Shin}
\email{shinkj@hyundai.com}
\affiliation{
 Materials Research \& Engineering Center, CTO Division, Hyundai Motor Company, Uiwang 16082, Republic of Korea
}

\author{Woomin Kyoung}
\affiliation{ 
  Materials Research \& Engineering Center, CTO Division, Hyundai Motor Company, Uiwang 16082, Republic of Korea
}

\author{Seunghyo Noh}
\affiliation{ 
  Materials Research \& Engineering Center, CTO Division, Hyundai Motor Company, Uiwang 16082, Republic of Korea
}

\author{Young Min Rhee}
\affiliation{Department of Chemistry, KAIST, Daejeon, 34141, Republic of Korea}

\author{Kyungmin Kim}
\affiliation{Department of Chemistry, KAIST, Daejeon, 34141, Republic of Korea}

\date{\today}
\maketitle

\onecolumngrid

\section{Construction of 4RDM}
In the implementation of the pair-VQE, the Hamiltonian is written as 
\begin{equation}
    \begin{split}
         H&=\sum_{pq}{g_{pq}\left(a^\dagger_{p\alpha}a_{q\alpha}+a^\dagger_{p\beta}a_{q\beta}\right)} \\
         &+\sum_{pqrs}{(pq|rs)\left(\frac{1}{2}a^\dagger_{p\alpha}a_{q\alpha}a^\dagger_{r\alpha}a_{s\alpha}+\frac{1}{2}a^\dagger_{p\beta}a_{q\beta}a^\dagger_{r\beta}a_{s\beta}\right)} \\
         &+\sum_{pqrs}{(pq|rs)a^\dagger_{p\alpha}a_{q\alpha}a^\dagger_{r\beta}a_{s\beta}} \\
    \end{split}
\end{equation}
in which $g$ is the modified one-electron integrals
\begin{equation}
    g_{pq}=h_{pq}-\frac{1}{2}\sum_r{(pr|rq)}
\end{equation}
and $(pq|rs)$ is the usual two-electron coulomb integrals in $(11|22)$ order. 

Therefore, for 4-RDMs, there are three different categories.
\begin{equation}
    \begin{split}
        \left<\Psi^{(0)}|a^\dagger_ua_va^\dagger_xa_y a^\dagger_{p\alpha}a_{q\alpha}a^\dagger_{r\alpha}a_{s\alpha}|\Psi^{(0)}\right> \\
        \left<\Psi^{(0)}|a^\dagger_ua_va^\dagger_xa_y a_{p\beta}^\dagger a_{q\beta}a^\dagger_{r\beta}a_{s\beta}|\Psi^{(0)}\right> \\
        \left<\Psi^{(0)}|a^\dagger_ua_va^\dagger_xa_y a^\dagger_{p\alpha}a_{q\alpha}a^\dagger_{r\beta}a_{s\beta}|\Psi^{(0)}\right> \\
    \end{split}
\end{equation}

For each category, there are three sub-categories, which corresponds to $\alpha\alpha$, $\beta\beta$, and $\alpha\beta$ excitations. For example, 
\begin{equation}
    \begin{split}
        \left<\Psi^{(0)}|a^\dagger_{u\alpha}a_{v\alpha}a^\dagger_{x\alpha}a_{y\alpha} a^\dagger_{p\alpha}a_{q\alpha}a^\dagger_{r\alpha}a_{s\alpha}|\Psi^{(0)}\right> (x<u, y<v, u\neq v,x\neq y,u\neq y,v\neq x) \\
        \left<\Psi^{(0)}|a^\dagger_{u\beta}a_{v\beta}a^\dagger_{x\beta}a_{y\beta} a^\dagger_{p\alpha}a_{q\alpha}a^\dagger_{r\alpha}a_{s\alpha}|\Psi^{(0)}\right> (x<u, y<v, u\neq v,x\neq y,u\neq y,v\neq x) \\
        \left<\Psi^{(0)}|a^\dagger_{u\alpha}a_{v\alpha}a^\dagger_{x\beta}a_{y\beta} a^\dagger_{p\alpha}a_{q\alpha}a^\dagger_{r\alpha}a_{s\alpha}|\Psi^{(0)}\right> (u\neq v,x\neq y, x\neq u || y\neq v) \\
    \end{split}
\end{equation}

As one could see, the first term can only contain 4 number operators to be non-zero. The second term can only contain 2 pair-double excitation operators. The third term can contain 2 number operators and one pair-double excitation operator. 

Take the first term for example, there are only 4 possible cases for this term to be non-zero. 
\begin{equation}
    \begin{split}
        u=q,v=p,x=s,y=r \\
        u=s,v=p,x=q,y=r \\
        u=q,v=r,x=s,y=p \\
        u=s,v=r,x=q,y=p \\
    \end{split}
\end{equation}

For second term, the possible cases are
\begin{equation}
    \begin{split}
        u=p,v=q,x=r,y=s \\
        u=r,v=q,x=p,y=s \\
        u=p,v=s,x=r,y=q \\
        u=r,v=q,x=p,y=s \\
    \end{split}
\end{equation}

Here we show how to build the 4RDM constructively.
The problem is defined as the following.
\begin{align}
    \left<\Psi^{(0)}|a^\dagger_ua_va^\dagger_xa_y a^\dagger_{p}a_{q}a^\dagger_{r}a_{s}|\Psi^{(0)}\right>
\end{align}
where $u,v,x,y,p,q,r,s \in \{0,1,\cdots,2n-1\}$ indicating spin orbitals where $n$ is the number of molecular orbitals.
The indices $u,v,x,y$ has the following additional constraints:
\begin{enumerate}
    \item $(u,v)$ and $(x,y)$ must belongs to the same spin respectively
    \item $u>x$, $v>y$
    \item As generalized excitation term, $u,v,x,y$ must not belong to same spin orbital, but they can be in the same molecular orbital of different spins
    \item $a^\dagger_ua_va^\dagger_xa_y$ must break the pair symmetry 
    \item $a^\dagger_ua_va^\dagger_xa_y a^\dagger_{p}a_{q}a^\dagger_{r}a_{s}$ must conserve the pair symmetry
\end{enumerate}

\begin{figure}[h]
\includegraphics[width=16cm]{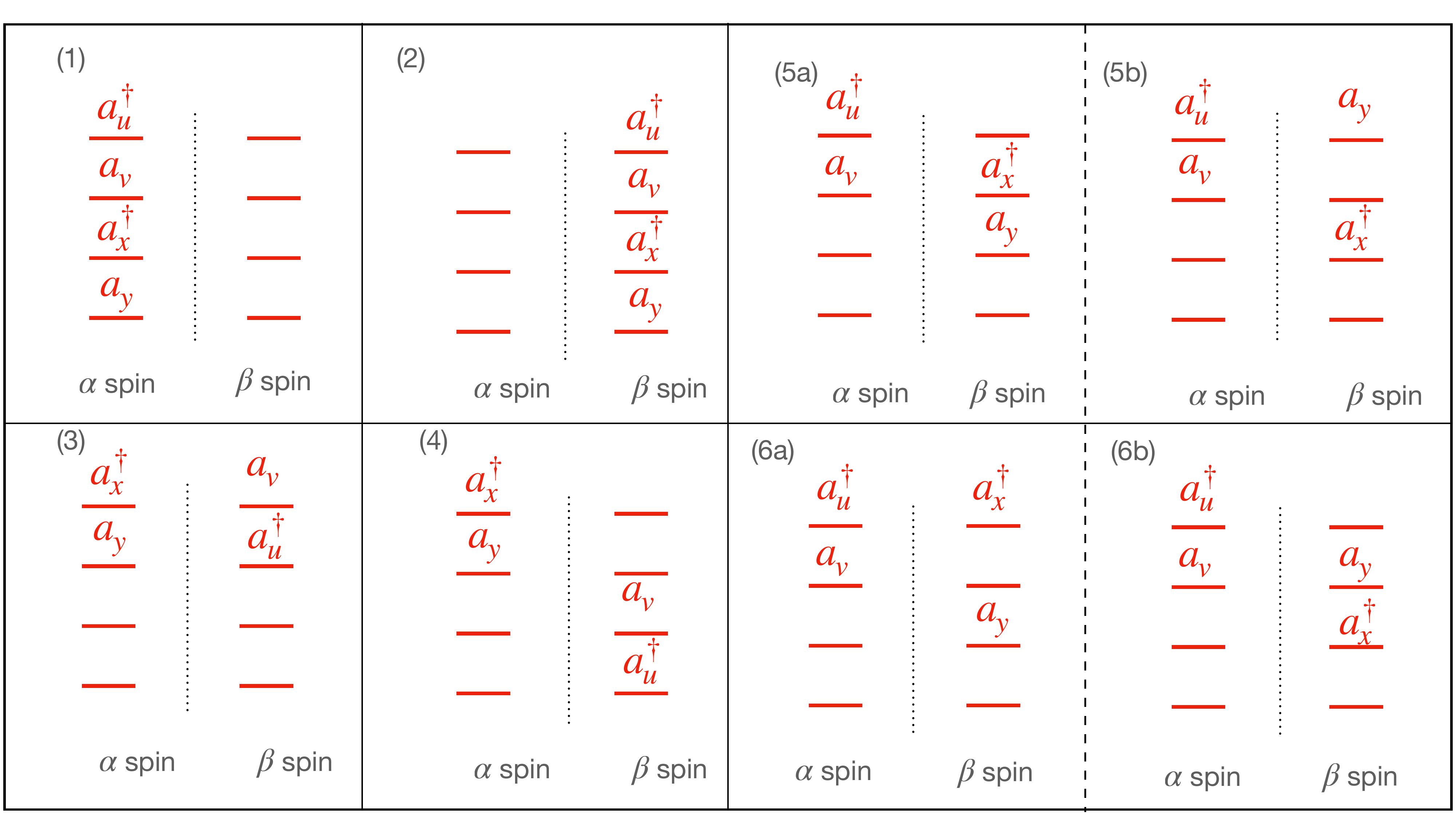}
\caption{All possible index combinations for $(uvxy)$ that breaks paired symmetry. 
The four levels corresponds to arbitrarily chosen four molecular orbitals among all possible molecular orbitals.
The two bars on the same level corresponds to the two spin orbitals of the same molecular orbital. 
The spin types of $(uvxy)$ are described as the following.
(1) $\alpha \alpha \alpha \alpha$ on different molecular orbitals. 
(2) $\beta \beta \beta \beta$ on different molecular orbitals.
(3) $\alpha \alpha \beta \beta$ on two molecular orbitals.
(4)  $\alpha \alpha \beta \beta$ on four different molecular orbitals.
(5) $\alpha \alpha \beta \beta$ on three molecular orbitals with (5a) annihilation-creation operators share the same molecular orbital or (5b) creation-annihilation operators share the same molecular orbital.
(6) $\alpha \alpha \beta \beta$ on three molecular orbitals with (6a) two creation operators share the same molecular orbital or (6b) two annihilation operators share the same molecular orbital.
}
\label{fig:uvxy}
\end{figure}

Since for each valid tuple of $(uvxy)$, there must be at least one valid $(pqrs)$ but not vise versa, it is more efficient to enumerate all valid $(uvxy)$ and derive its corresponding valid $(pqrs)$.
All valid types of $(uvxy)$ are shown in Fig.~\ref{fig:uvxy} .
For example, for (1) in Fig.~\ref{fig:uvxy}, it means out of all molecular orbitals, choose four of them and assign the $\alpha$ spin part of the molecular orbital to $(uvxy)$. 
Notice, however, when enumerate all possible $(uvxy)$, it still needs to satisfy the additional condition that $x<u, y<v$.

For each of the above cases, we need to derive a matching $(pqrs)$ so that the combined eight indices sill preserve the symmetry.
We showed all possible combinations of $(uvxy)$ and $(pqrs)$ in Fig.~2.

\begin{figure}[h]
\label{fig:uvxy-pqrs}
\includegraphics[width=16cm]{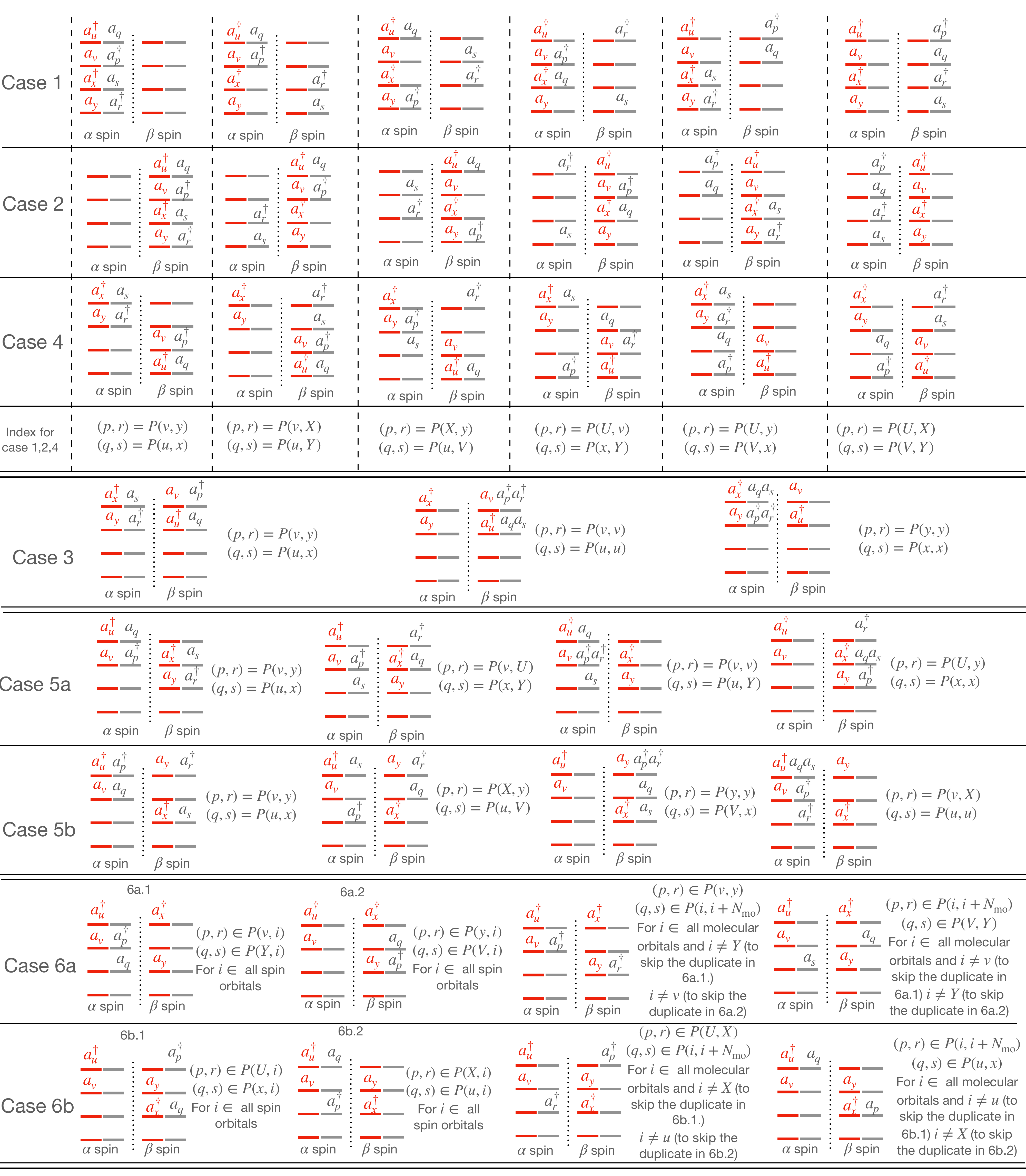}
\caption{
The figure lists the rules to generate valid indices for 
$\left<\Psi^{(0)}|
{\color{red}a^\dagger_ua_va^\dagger_xa_y}
{a^\dagger_pa_qa^\dagger_ra_s}
|\Psi^{(0)}\right>$ for a given ${\color{red}a_u^\dagger a_v a_x^\dagger a_y }$. 
All indices are spin orbital indices.
$P(\cdot,\cdot)$ represents the permutation function to generate valid indices and is defined as
$P(x,y) = 
\left\{ 
\begin{matrix}
\{(x,y),(y,x)\} &\mathrm{\ if\ } x\ne y\\
\{(x,x)\} &\mathrm{\ if\ } x=y
\end{matrix} 
\right.
$.
The capitalization of the letter indicates a index of opposite spin.
For example, $U=u+N_\mathrm{mo}$ if $u$ is $\alpha$ spin and  $U=u-N_\mathrm{mo}$ if $u$ is $\beta$ spin, where $N_\mathrm{mo}$ is the number of molecular orbitals.
For example, for all three cases (case 1, 2 and 4) in the first column, the $(p,r)$ should be $P(v,y)=\{(v,y),(y,v)\}$ which means $p=v,r=y$ or $p=y,r=v$, while $(q,s)$ should be $P(u,x)=\{(u,x),(x,y)\}$. 
So the valid operators are 
${\color{red}a_u^\dagger a_v a_x^\dagger a_y } a_v^\dagger a_u a_y^\dagger a_x $,
${\color{red}a_u^\dagger a_v a_x^\dagger a_y } a_v^\dagger a_x a_y^\dagger a_u $,
${\color{red}a_u^\dagger a_v a_x^\dagger a_y } a_y^\dagger a_u a_v^\dagger a_x $,
${\color{red}a_u^\dagger a_v a_x^\dagger a_y } a_y^\dagger a_x a_v^\dagger a_u $.
}
\end{figure}

\clearpage
\section{Efficient post-processing}
In this section we explain another major effort to improve the speed for evaluating the expectation value of second-quantized Hamiltonians that satisfy pair-symmetry.
Since the second quantized operator $O_J$ to be measured satisfies the pair symmetry, it can always be partitioned as 
\begin{align}
O_J = \prod_{jk} a^\dagger_j a_k
\rightarrow (-1)^b \prod_{jk}{a_{j \alpha}^\dagger a_{j \beta}^\dagger
a_{k \alpha} a_{k \beta}}
\prod_{l}a_l^\dagger a_l 
\prod_{m}  a_m a_m^\dagger 
\end{align}
that is, paired excitation parts, and number operator parts.
We then expand the fermionic operators with the following transformation
\begin{align}
    a_{j \alpha}^\dagger a_{j \beta}^\dagger & = \frac{1}{2}(X_j-iY_j) \\
    a_{j \alpha}  a_{j \beta} & = 
-\frac{1}{2}(X_j+iY_j) \\
a_l^\dagger a_l &= \frac{1}{2}(I-Z_l)\\
a_l a_l^\dagger &= \frac{1}{2}(I+Z_l)
\end{align}
and take only the real part.
So each of the 
\begin{align}
O_J = \prod_{jk} a^\dagger_j a_k
\mapsto
\sum_j h_j \otimes_k \sigma_v^{(jk)} = \sum_j h_j P_j
\end{align}
where $\sigma_v \in \{I,X,Y,Z\}$ is one of the Pauli operators and $P_j$ denotes a Pauli string.
As a result, in order to obtain $\langle O_J \rangle$, we only needs to obtain 
$\langle P_j \rangle$.

Notice there will be many $P_j$ that are duplicates from different $O_J$.
To avoid duplicate measurement and post-processing, we first collect all unique $P_j$  from all $\{O_J\}$ of interest to form a set $\{P_j\}$, obtain $\{ \langle P_j \rangle \}$ and store it into a hash table (such as dictionary in Python).
Then to recover the $\langle O_J \rangle$, we only need to retrieve the $\langle P_j \rangle$ from $\{ \langle P_j \rangle \}$ and then sum them up with the weight $h_j$ in each $O_J$.
It is worth noting that, when measure the expectation value for $\{P_j\}$, we can measure several Paul strings simultaneously without the increase of the number of two-qubit gates as long as all Pauli strings to be measured simultaneously are mutually qubit-wise-commuting.
In practice we use the qiskit qubit-wise-commuting function.

\clearpage
\bibliography{Journal_Short_Name, main}